\documentclass[prx,twocolumn,showpacs,amsmath,amssymb,superscriptaddress]{revtex4-2}
\usepackage{graphicx}
\usepackage{dcolumn}
\usepackage{bm}
\usepackage{hyperref}
\hypersetup{colorlinks,citecolor=blue, filecolor=blue, linkcolor=blue , urlcolor=blue}

\begin{document}

\title{Manipulation of localized excitons in CrPS$_4$ by temperature and magnetic field}

\author{Dipankar Jana}
    \email{jana.d02@nus.edu.sg}
    \affiliation{Institute for Functional Intelligent Materials, National University of Singapore, 117544, Singapore}

\author{Swagata Acharya}
    \affiliation{National Laboratory of the Rockies, Golden, CO, 80401 USA}

\author{Amit Pawbake}
    \affiliation{Laboratoire National des Champs Magn\'etiques Intenses, LNCMI-EMFL, CNRS UPR3228,Univ. Grenoble Alpes, Univ. Toulouse, Univ. Toulouse 3, INSA-T, Grenoble and Toulouse, France}

\author{Dmitrii Litvinov}
    \affiliation{Institute for Functional Intelligent Materials, National University of Singapore, 117544, Singapore}
    \affiliation{Department of Materials Science and Engineering, National University of Singapore, 117575, Singapore}

\author{Aljoscha Soll}
    \affiliation{Department of Inorganic Chemistry, University of Chemistry and Technology Prague, 16628 Prague, Czech Republic}

\author{Zdenek Sofer}
    \affiliation{Department of Inorganic Chemistry, University of Chemistry and Technology Prague, 16628 Prague, Czech Republic}

\author{Clement Faugeras}
    \affiliation{Laboratoire National des Champs Magn\'etiques Intenses, LNCMI-EMFL, CNRS UPR3228,Univ. Grenoble Alpes, Univ. Toulouse, Univ. Toulouse 3, INSA-T, Grenoble and Toulouse, France}

\author{Dimitar~Pashov}
    \affiliation{King’s College London, Theory and Simulation of Condensed Matter, The Strand, WC2R 2LS London, UK}   

\author{Mark van Schilfgaarde}
    \affiliation{National Laboratory of the Rockies, Golden, CO, 80401 USA}

\author{Kostya S. Novoselov}
    \email{kostya@nus.edu.sg}
    \affiliation{Institute for Functional Intelligent Materials, National University of Singapore, 117544, Singapore}
    \affiliation{Department of Materials Science and Engineering, National University of Singapore, 117575, Singapore}

\author{Marek~Potemski}
    \affiliation{Laboratoire National des Champs Magn\'etiques Intenses, LNCMI-EMFL, CNRS UPR3228,Univ. Grenoble Alpes, Univ. Toulouse, Univ. Toulouse 3, INSA-T, Grenoble and Toulouse, France}
     \affiliation{CENTERA, CEZAMAT, Warsaw University of Technology, 02-822 Warsaw, Poland} 
    \affiliation{Institute of High Pressure Physics, PAS, Warsaw, PL-01-142 Poland} 

\author{Maciej~Koperski}
    \email{msemaci@nus.edu.sg}
    \affiliation{Institute for Functional Intelligent Materials, National University of Singapore, 117544, Singapore}
    \affiliation{Department of Materials Science and Engineering, National University of Singapore, 117575, Singapore}

\begin{abstract}

Layered van der Waals magnetic semiconductors provide a versatile platform for exploring excitonic phenomena intertwined with spin and lattice degrees of freedom, enabling excitons to act as sensitive probes of magnetic order. CrPS$_4$ is a layered antiferromagnetic semiconductor that hosts rich excitonic features whose microscopic origin and connection to magnetic ordering remain incompletely understood. Here, we investigate the electronic and excitonic properties of bulk CrPS$_4$ using a combination of many-body perturbation theory, dynamical mean-field theory, and photoluminescence-based experiments. Our calculations establish CrPS$_4$ as a direct-gap semiconductor with a bandgap of 2.48~eV in the antiferromagnetic phase. Several sub-bandgap excitonic transitions are predicted by theory, comprising multiple spin-allowed excitons and an additional spin-flip excitation, predominantly localized on the Cr$^{3+}$ ions. Temperature- and magnetic-field-dependent optical measurements reveal thermally driven exciton redistribution among localized states and identify characteristic energy shifts that provide clear optical signatures of magnetic phase transitions in CrPS$_4$. These results provide new insights into the excitonic transitions of antiferromagnets and suggest potential routes for all-optical sensing and light-driven control of their magnetic order.

\end{abstract}

\maketitle


\section{Introduction}

Two-dimensional van der Waals magnets have emerged as a powerful platform for the generation, transport, and detection of spin waves, with promising applications in spintronics and quantum information technologies~\cite{2D_magnets_review, de2023long, wang2022, chen2021electrically, qiu2025hallmarks, rezende2016diffusive}. Beyond magnonic functionalities, these materials provide a compelling arena for exploring the interplay between magnetism and strong light–matter interactions. Recent studies have predominantly focused on van der Waals antiferromagnets, where a pronounced coupling between long-range magnetic order and highly localized excitons has been observed below the N\'eel temperature~\cite{Kang2020, wilson2021interlayer}. The magnetic order that dictates the spin alignment can be modified by external factors such as temperature, magnetic fields, or strain, which in turn influence both the band structure and excitonic properties of the material. Conversely, the optical response driven by the excitonic transitions can provide a deep insight into the magnetic ordering~\cite{Dipankar2023, jana2025deconstruction, pawbake2022, jana2025two, CrX3_optical_spin_pump, CrBr3_topological_textures, CrBr3_skyrmion_dynamics, CrBr3_resonant_Raman, CrBr3_bubble, CrBr3_domain_correlations, CrCl3_proximity_effects}, and there is the intriguing possibility that electronic excitation to certain excitonic states could influence the magnetic order of these materials~\cite{ilyas2024terahertz, xie2025high}.

CrPS$_4$ is a two-dimensional A-type antiferromagnet from the MPX$_n$ family, where M is a transition metal atom such as Mn, Ni, Fe, Co, or Cr; X is a chalcogen atom such as S or Se; and $n$ = 3, 4. It crystallizes in a monoclinic structure belonging to the $C2$ space group~\cite{Diehl1977}. The magnetic properties of CrPS$_4$~\cite{gu2019, multian2025brightened, riesner2022temperature} have been studied extensively by several techniques~\cite{peng2020magnetic, Calder2020, peng2022controlling, Pei2016, huang2023layer, wu2023gate, ho2025imaging, son2021air}. The stable magnetic order, below the N\'eel temperature of 38~K, is shown schematically in Fig.~\ref{fig:Fig1}a~\cite{Momma2011}. Within each layer, the Cr$^{3+}$ magnetic moments align ferromagnetically along the crystallographic $c$-axis, with a small tilt of approximately 12.5$^{\circ}$ toward the a-axis, while moments in adjacent layers are coupled antiferromagnetically~\cite{bud2021magnetic, Calder2020, peng2020magnetic}. The combined effects of interlayer exchange coupling and biaxial magnetic anisotropy give rise to two distinct magnon gaps at approximately 13~GHz and 25~GHz~\cite{li2023ultrastrong, freeman2025tunable}. Applying an external magnetic field along the magnetic axes counteracts the antiferromagnetic exchange coupling and magnetic anisotropy, driving successive spin reorientation transitions at the spin-flop field of approximately 1~T and the critical saturation field of about 8~T, respectively \cite{wu2023gate, bud2021magnetic, peng2020magnetic, jana2026spin}. 

Besides its magnetic properties, CrPS$_4$ is a semiconductor whose band gap characteristics have been debated due to the emergence of electronic correlations. Different levels of theory predict vastly different band gap values, while the experimental validations are challenging due to the rich and complex spectroscopic responses. This results in contradicting reports claiming values of the band gap equal to 1.4~eV or 2.4~eV~\cite{gu2019, lee2017structural, louisy1978physical, budniak2020exfoliated, multian2025brightened, bo2023magnetic, alcantara2023parameter}.  The sub-bandgap optical transitions are attributed to localized $d$–$d$ excitations of the Cr$^{3+}$ ions~\cite{riesner2022temperature, multian2025brightened, kim2022photoluminescence, gu2019}, which are interpreted within the framework of the Tanabe–Sugano (T-S) diagram~\cite{Tanabe1954} for a $d^3$ electron configuration. Of particular interest are the spectrally narrow emission lines, which are commonly assigned to spin-flip transitions ($^2$E, $^2$T$_1$, and $^2$T$_2$  $\rightarrow$ $^4$A$_2$) but have not been conclusively shown to correlate with the magnetic ordering of the material~\cite{riesner2022temperature, multian2025brightened, gu2019, hu2025magnetic}. In contrast to spin-flip transitions reported in other MPX$_3$ antiferromagnets~\cite{Dipankar2023, Gnatchenko2011, stager1963zeeman, van1967optical, russell1966zeeman}, these features neither diminish above the N\'eel temperature nor exhibit a Zeeman shift, questioning their assignment within the T-S framework.

In this work, we systematically examine the electronic properties of bulk CrPS$_4$ using many-body perturbation theory (MBPT), dynamical mean-field theory (DMFT), and experiments based on photoluminescence (PL) spectroscopy, directly unraveling the role of electronic correlations in the formation of excitonic states and their coupling to magnetic order. Our results demonstrate that CrPS$_4$ is a direct-gap semiconductor with a bandgap of 2.48~eV in the antiferromagnetic phase. Several spin-allowed transitions, together with an additional spin-flip transition, are calculated in the vicinity of 1.4~eV, with the lowest-energy spin-allowed transition being optically dark. The spin-allowed nature of the dominant transitions is confirmed by the absence of a Zeeman shift, while their characteristic energy shifts with temperature and magnetic field provide a clear signature of the magnetic phase transition points in CrPS$_4$. A simultaneous enhancement of PL intensity and decay time is observed and attributed to thermally activated exciton transfer from a magnetically induced localized state to an optically bright state.

\section{Results and discussion}

\subsection{Electronic band structure of $\mathrm{CrPS_4}$}

\begin{figure*}[htp]
    \includegraphics[width=17cm]{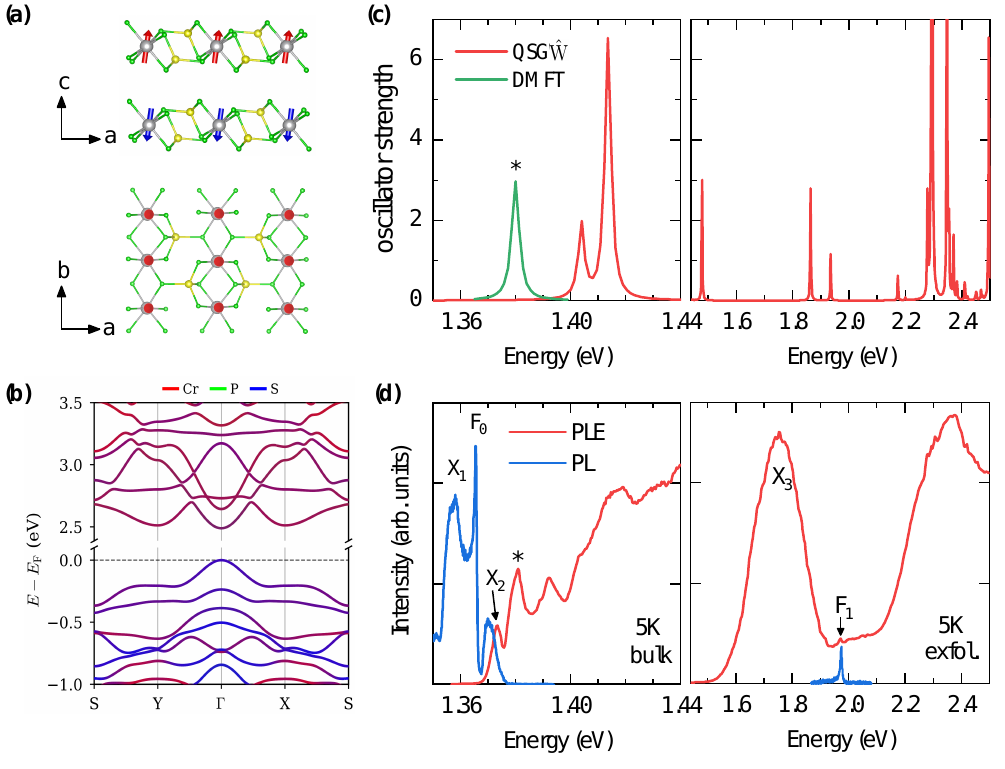}
	\caption{\textbf{(a)} Schematic crystallographic and magnetic structure of CrPS$_4$ in the antiferromagnetic phase. Gray, green, and yellow spheres denote chromium, sulfur, and phosphorus atoms, respectively. Images generated using the VESTA software package~\cite{Momma2011}. \textbf{(b)} Electronic band structure of CrPS$_4$ in the AFM phase. The Fermi level is set to zero. The red- and blue-colored bands correspond to Cr and S ions, respectively. \textbf{(c)} Optical absorption spectra of CrPS$_4$ computed using the $\mathrm{QSG\hat{W}}$ method for the spin-allowed transitions and the DMFT method for the spin-flip transition. An artificial linewidth of 3~meV broadens all features. \textbf{(d)} Low-temperature (5~K) PL and PLE spectra of CrPS$_4$, measured on a bulk (thickness~$\approx 500~\mu$m) and an exfoliated flake (thickness~$\approx 10-20~\mu$m). The PLE spectrum is obtained by collecting the integrated PL intensity of the low-energy tail ($\le$ 1.3~eV). The feature marked by ‘*’ corresponds to a spin-flip transition ($\Delta m_s = 1$), whereas all other features are spin-allowed ($\Delta m_s = 0$).}
	\label{fig:Fig1}
\end{figure*}

\begin{table}[]
\centering
\caption{Low-energy spin-allowed excitons and their corresponding oscillator strengths predicted by quasi-particle self-consistent $G\hat{W}$ approach.}
\begin{tabular}{ll}
\hline
Energy (eV) & Oscillator strength \\ \hline
1.379 & 3.8x10$^{-7}$ \\
1.404 & 1.8 \\
1.414 & 6.5 \\
1.428 & 2.0x10$^{-6}$ \\
1.440 & 1.7x10$^{-5}$ \\
1.479 & 3.1 \\
\hline
\end{tabular}
\label{Table:1}
\end{table}

The electronic structure of CrPS$_4$ bulk crystal is computed with a self-consistent \textit{ab initio} MBPT approach, quasi-particle self-consistent \textit{GW} (QS$GW$) \cite{qsgw,questaal_paper} and QS$G\hat{W}$ approaches~\cite{Cunningham2023}. QS$G\hat{W}$~\cite{Cunningham2023} is a self-consistent extension of QS$GW$ where the electronic structure is computed in the presence of the ladder electron-hole vertex corrections, which become crucial in magnetic systems with partially filled $d$ and $f$ states~\cite{acharyaTheoryColorsStrongly2023,acharya2021electronic,pashov2025quasiparticle,watson2024giant}, which are narrow and significantly screened. The corresponding $QSG\hat{W}$ band structures, shown in Fig.~\ref{fig:Fig1}b, reveals a bulk electronic bandgap ($E_g$) of 2.48~eV in the antiferromagnetic phase. The exciton spectra are computed with $\mathrm{QS}G\hat{W}$ framework and exact computation of higher order charge-charge correlators within DMFT, using an exact-diagonalization (ED) impurity solver~\cite{acharyaTheoryColorsStrongly2023}. $\mathrm{QS}G\hat{W}$ is limited to spin-allowed excitonic transitions, but the nonperturbative, locally exact impurity vertex in ED-DMFT produces the spin-flip atomic multiplet transitions missing from MBPT but localized solely to the Cr$^{3+}$ ion. For CrPS$_{4}$, $\mathrm{QS}G\hat{W}$ predicts several spin-allowed excitons around 1.4~eV, along with additional transitions at 1.87~eV and 1.93~eV, whereas ED-DMFT captures a spin-flip transition at 1.38~eV (missing from $\mathrm{QS}G\hat{W}$), as shown in Fig.~\ref{fig:Fig1}c and summarized in Table~\ref{Table:1} with the corresponding oscillator strength.

These electronic transitions are examined using low-temperature PL and PL excitation (PLE) measurements. Figure~\ref{fig:Fig1}d shows the corresponding spectra measured on a bulk (thickness~$\approx 500~\mu$m) and an exfoliated CrPS$_4$ flake (thickness~$\approx 10-20~\mu $m)  at 5~K. The PL spectrum is dominated by several narrow transitions (X$_1$ and F$_0$) near 1.36~eV and a high-energy transition (F$_1$) at 1.97~eV. The PL spectra remain essentially unchanged for the two thicknesses studied. The PLE spectrum is obtained by integrating the PL intensity of the low-energy tail ($\le$ 1.3~eV). The bulk sample is used to enhance the visibility of weak oscillator-strength features near 1.4~eV (X$_2$ and *), while the exfoliated sample is employed to investigate the bandgap. The PLE spectrum of the exfoliated flake reproduces an absorption edge in close agreement with the theoretically predicted band gap and the absorption spectrum (\textit{Supplemental Material (SM)}, Fig.~S1~\cite{SuppInfo}). In addition to the band-edge response, a broad feature (X$_3$) at 1.78~eV and F$_1$ are also observed in the PLE spectrum of the exfoliated flake. The pronounced broadening of the X$_3$ transition is attributed to multi-phonon–assisted processes, which are not considered in the theoretical calculations. The F$_0$ and F$_1$ transitions exhibit a Fano line shape arising from their interference with the relatively broad X$_1$ and X$_3$ transitions, respectively. 


\subsection{Temperature and magnetic manipulation of excitons}

\begin{figure*}[htp]
\centering
\includegraphics[width=17cm]{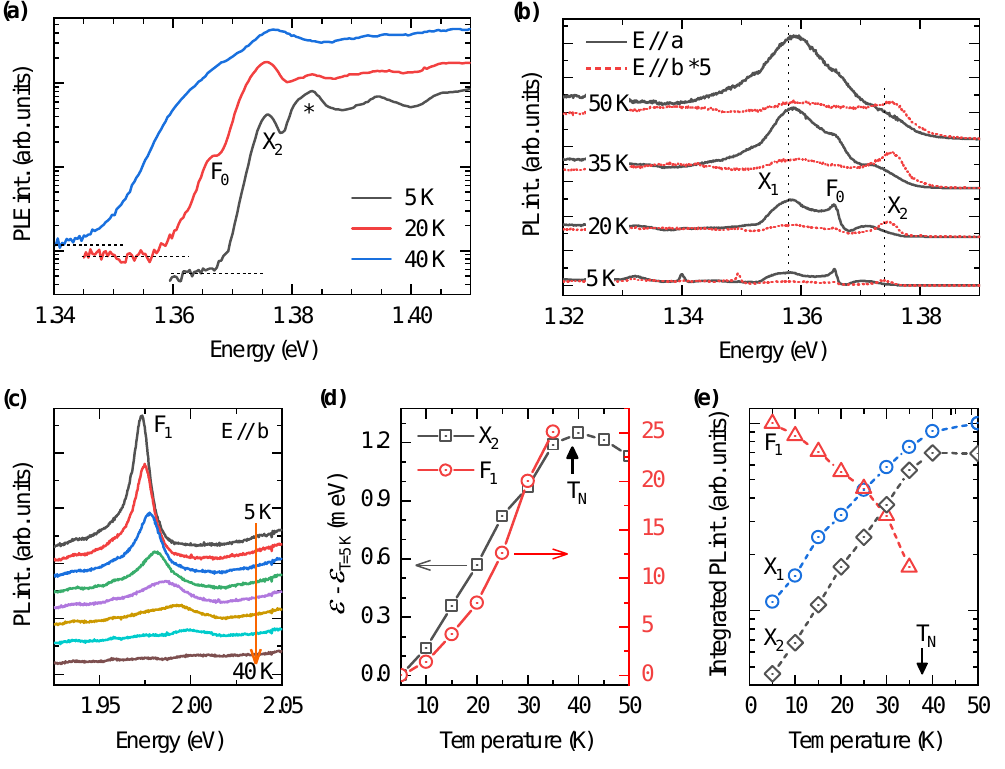}
\caption{\textbf{(a)} Unpolarized PLE spectra and \textbf{(a)} linear polarization resolved PL spectra of bulk CrPS$_4$ at some selected temperatures. Spectra are vertically shifted for clarity. PL spectra acquired with $E \parallel b$-axis are scaled by a factor of 5.  Vertical dotted lines serve as a guide to the temperature-dependent peak shift of the dominant transitions, while the horizontal dotted lines in the PLE plot correspond to the background level. \textbf{(c)} PL spectra of bulk CrPS$_4$ in the high-energy region, measured in the $E \parallel b$-axis configuration with 5~K temperature intervals. Temperature-dependent \textbf{(d)} peak shifts of the X$_2$ and F$_1$ transitions and \textbf{(e)} integrated intensities of the X$_1$, X$_2$, and F$_1$ transitions. The intensities of X$_2$ and F$_1$ are scaled. The vertical arrows indicate the N\'eel temperature (T$_N$=38~K).}
\label{fig:Fig2}
\end{figure*}

\begin{figure*}[htp]
\centering
\includegraphics[width=17cm]{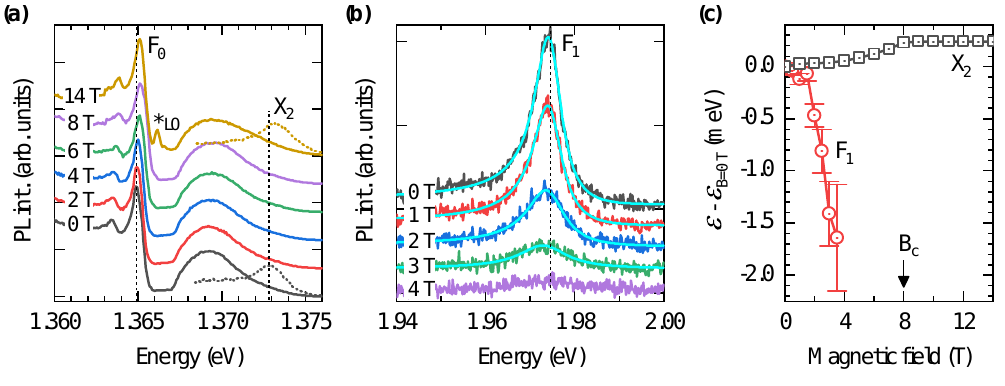}
\caption{Low-temperature PL spectra of bulk CrPS$_4$ as a function of a magnetic field applied perpendicular to the sample plane ($E \parallel c$-axis) in \textbf{(a)} low-energy and  \textbf{(b)} high-energy spectral windows. Vertical dashed lines serve as visual guides for field-dependent peak shift. '*$_\mathrm{LO}$' corresponds to the phonon replica of spin-flip transition '*'~\cite{jana2026spin}. \textbf{(c)} Magnetic field dependence of the peak energies of the X$_2$ and F$_1$ transitions. The vertical arrow indicates the critical saturation field ($B_c$=8~T) at which CrPS$_4$ undergoes a transition from canted AFM to a FM state.}
\label{fig:Fig3}
\end{figure*}

A key difference between our theoretical calculations and experimental results reported in the literature lies in the assignment of the optical transitions around 1.4~eV. Previous studies attribute the dominant features (X$_1$ and F$_0$) to the $^2$E, $^2$T$_1$ $\leftrightarrow$ $^4$A$_2$ transitions, which are interpreted as spin-flip transitions within the Tanabe–Sugano framework~\cite{multian2025brightened, kim2022photoluminescence, riesner2022temperature}. In contrast, our $\mathrm{QS}G\hat{W}$ calculations reveal that most transitions near 1.4~eV are spin-allowed. In the following, we demonstrate that these transitions do not follow the characteristic temperature or magnetic field evolution expected for spin-flip exciton as reported for other MPX$_n$ materials~\cite{jana2025deconstruction, Dipankar2023, Gnatchenko2011}.

Figure~\ref{fig:Fig2}a shows the unpolarized PLE spectra of bulk CrPS$_4$ measured by integrating the low-energy PL tail ($\le$~1.3~eV) at a few selected temperatures. The narrow resonance marked by '*' disappears above the N\'eel temperature and has been attributed to a spin-flip transition ($\Delta m_s$ = 1) between the excited $m'_s = 1/2$ and the ground $m_s = 3/2$ spin states of Cr$^{3+}$ ions~\cite{jana2026spin}. The observation is consistent with our DMFT calculations, which predict a spin-flip transition in CrPS$_4$ at 1.38~eV. The X$_2$ feature becomes more prominent, while F$_0$ and X$_1$ features are observed at elevated temperatures. The X$_1$ transition manifests as a low-energy tail of the PLE edge at higher temperatures rather than a distinct peak. The increase in the PLE intensity of the low-energy excitons reflects changes in the oscillator strength and the occupation of the corresponding excitonic states. A temperature-induced blueshift of the X$_2$ transition is observed in the PLE spectra.  This is further confirmed by temperature-dependent PL measurements. The linear polarization resolved PL spectra of bulk CrPS$_4$ at selected temperatures are shown in Fig.~\ref{fig:Fig2}b and Fig.~\ref{fig:Fig2}c. The dominant PL features, X$_1$ and F$_0$, are linearly polarized along the crystallographic a-axis ($E \parallel a$), while the X$_2$ feature is detected in $E \parallel b$-axis polarization. The emission energies of these excitons coincide with their corresponding excitation energies, indicating absence of a Stokes shift. All features near 1.4~eV exhibit a small but systematic blueshift, accompanied by an increase in PL yield as temperature increases. The F$_0$ transition shows a similar enhancement up to 20-25~K, beyond which its intensity progressively diminishes (\textit{SM}, Fig.~S2~\cite{SuppInfo}). The F$_1$ transition appears in $E \parallel b$-axis polarization and exhibits a comparatively stronger blueshift with increasing temperature, accompanied by a decrease in intensity.

Figures~\ref{fig:Fig2}d and \ref{fig:Fig2}e summarize the temperature dependence of the peak energies and integrated PL intensities of the PL features, respectively. The temperature-dependent blueshift of the 1.4~eV features persists up to the antiferromagnetic-to-paramagnetic phase transition temperature. Due to the pronounced broadening of the X$_1$ feature and the asymmetric lineshape of the F$_0$ transitions, reliable extraction of their peak shifts is not feasible. Nevertheless, the vertical dashed lines in Fig.~\ref{fig:Fig2}b indicate an apparent blueshift of the X$_1$ transition. The high-energy F$_1$ transition displays a more pronounced blueshift than the 1.4~eV excitons. However, it is no longer observed above T$_N$. The blueshift of the excitons is attributed to an increase in bandgap of CrPS$_4$ upon transitioning from the antiferromagnetic to the paramagnetic phase, as predicted by theoretical calculations (\textit{SM}, Fig.~S3~\cite{SuppInfo}). The observed blueshift in exciton energy with increasing temperature beyond the magnetic ordering point is a definitive signature of magneto-optical coupling. Similar trends have been reported in several other low-dimensional magnets~\cite{shao2024exciton,datta2025magnon,ruta2023hyperbolic,jana2025deconstruction,acharya2025spin}, and their microscopic origin was systematically explored in the context of \mbox{CrSBr} in previous combined theoretical and experimental studies. Across the magnetic transition, two competing energy renormalizations must be considered. The conventional temperature-driven effects (thermal expansion and electron-phonon renormalization) tend to reduce the gap~\cite{vina1984temperature}, but in CrPS$_4$ the magnetic-disorder contribution is larger near $T_N$: when the ferromagnetic chains lose long-range order in the paramagnetic phase, spin-conserving hopping is reduced, and correlation-induced localization is enhanced. Within our self-consistent, parameter-free QSG$\hat{W}$ framework, this produces a substantial increase of the quasiparticle band gap, from $2.48$~eV in the antiferromagnetic phase to $2.66$~eV in the paramagnetic phase (\textit{SM}, Fig.~S3~\cite{SuppInfo}), while simultaneously increasing the exciton binding energy through stronger real-space confinement (smaller effective Bohr radius). The resulting exciton shift is therefore the difference $\Delta \mathcal{E}_X=\Delta \mathcal{E}_g-\Delta \mathcal{E}_b$, which is much smaller than either term individually. This observation justifies why low-energy excitons exhibit only modest net shifts despite large underlying electronic renormalizations. In parallel, enhanced paramagnetic-phase localization increases onsite $d$-$d$ character at the expense of intersite $d$-$d$/$d$-$p$ components, enforcing the Laporte parity constraint more strictly and reducing oscillator strength (darker excitons), consistent with Fig.~S3 in the \textit{SM}~\cite{SuppInfo}. This provides a natural explanation for the reduction in oscillator strength of the '*', F$_0$, and F$_1$ features when approaching T$_N$ as shown in Fig.~\ref{fig:Fig2}a,b, and c. In contrast, the increase in the intensities of the X$_1$ and X$_2$ features up to 50~K cannot be explained by exciton wavefunction localization, likely due to their strong coupling with phonons, as evidenced by the presence of multiple phonon replicas (\textit{SM}, Fig.~S2~\cite{SuppInfo}). For these strongly localized, nearly onsite $d$-$d$ excitons, phonons can renormalize the gap and reshape spectral weight through Huang--Rhys/Franck--Condon vibronic effects, including linewidth and sideband redistribution (e.g., broad $X_3$-like features), which are not included in the present electronic calculation. Thus, the exact experimental blue/red shift reflects a coupled correlation-plus-phonon problem. The present work unambiguously isolates the electronic-correlation contribution in a parameter-free manner.

Magneto-PL spectroscopy is performed to investigate the coupling between excitons and magnetic ordering. Figures~\ref{fig:Fig3}a~and~\ref{fig:Fig3}b display the PL spectra near 1.4~eV and 1.97~eV, respectively, measured as a function of a magnetic field applied perpendicular to the sample plane ($B\parallel c$-axis). The low-energy excitons exhibit a progressive blueshift with increasing magnetic field up to 8~T, above which their energies become essentially field independent. In contrast, the high-energy exciton (F$_1$) displays a more pronounced redshift with increasing field but disappears at a threshold field of approximately 4~T. The corresponding magnetic-field-dependent peak energies of the X$_2$, as a representative of low-energy excitons, and F$_1$ excitons are summarized in Fig.~\ref{fig:Fig3}c.

Several conclusions can be drawn from these observations. First, the low-energy excitons exhibit a weak magnetic field-dependent blueshift up to the critical saturation field (B$_c\approx$~8~T~\cite{bud2021magnetic, jana2026spin}), at which CrPS$_4$ undergoes a transition to a ferromagnetically aligned spin state, establishing a coupling between the excitons and the magnetic order. While the origin of these opposite field dependencies of X$_3$ and F$_1$ transitions is not fully understood, it can be attributed to spin-orientation-dependent modifications of the bandgap and the exciton binding energy~\cite{wilson2021interlayer} (\textit{SM}, Fig.~S3~\cite{SuppInfo}). Second, unlike the spin-flip transition, none of these observed transitions show a spin-ordering-dependent Zeeman shift under an applied magnetic field. In this magnetically ordered material, the spin quantization axis is well defined even in the absence of an external magnetic field. Consequently, the Zeeman shift of an excitonic state depends on the relative orientation between the applied field and the spin quantization axis, the spin projection quantum number, and the g-factors of the ground and excited states. The exciton energy in the presence of an external magnetic field can therefore be expressed as~\cite{sell1966magnetic},

\begin{equation} \label{eq:2}
\mathcal{E} = \mathcal{E}_{B=0~T} + (g{'} m{'}_s - g m_s)\mu_B \textbf{B}\cdot \hat{\textbf{S}}
\end{equation}

where the primed (unprimed) quantities $g$ and $m_s$  denote the $g$-factor and spin projection quantum number of the excited (ground) state, respectively, and $\hat{\textbf{S}}$ represents the spin direction. Application of a magnetic field along the spin quantization axis lifts the degeneracy between the two antiferromagnetic sublattices, while a distinct magnetic-field dependence emerges in the canted and ferromagnetic phases~\cite{jana2025deconstruction}. In contrast, no such shift or splitting is observed for either the low- or high-energy excitons in CrPS$_4$. This absence of a Zeeman response implies that the term $(g' m'_s - g m_s)$ is zero for these excitons, indicating that the ground and excited states possess identical $g$-factors ($g'=g$) and spin projection quantum numbers ($m'_s=m_s$). This conclusion is consistent with our calculations, which identify these excitons as spin-allowed transitions.

\subsection{Exciton dynamics}

\begin{figure}[htp]
\centering
\includegraphics[width=8.4cm]{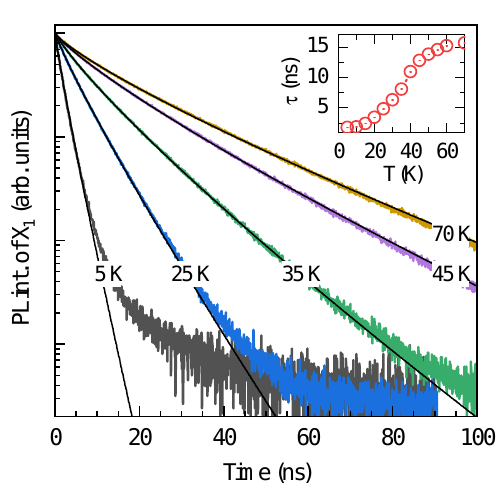}
\caption{Temporal decay profile of the X$_1$ PL intensity at some selected temperatures. All decay profiles are normalized to their peak intensity. The black lines denote fits to the data using a single stretched-exponential decay function. The inset shows the estimated temperature dependence of the decay time of X$_1$ transition.}
\label{fig:Fig4}
\end{figure}

To gain insight into the redistribution of excitons among the electronic states, we investigated the temperature dependence of their decay dynamics. Figure~\ref{fig:Fig4} shows the decay traces of the X$_1$ transition, measured in $E \parallel a$-axis polarization, at a few selected temperatures. The decay dynamics of the other low-energy transitions are nearly identical (\textit{SM}, Fig.~S4~\cite{SuppInfo}). At 5~K, the decay trace exhibits multiexponential behavior and is quantitatively described by a stretched exponential function, $I(t)=I_0 \exp[-(t/\tau)^\beta]$. The simulated solid curve provides an excellent fit to the experimental data with $\beta \approx 0.9$, accurately reproducing the decay over more than two orders of magnitude in intensity. As the temperature increases, the decay of the X$_1$ transition progressively slows down up to 70~K. The resulting temperature dependence of the X$_1$ decay time is summarized in the inset of Fig.~\ref{fig:Fig4}. The measured PL decay time of an excitonic state reflects its intrinsic radiative and nonradiative recombination channels, thermal coupling to nearby states, and the activation energies governing inter-state population transfer. An increase in exciton decay time with temperature has been reported to arise either from thermal redistribution of excitons in momentum space or from coupling to nearby dark excitonic states~\cite{feldmann1987linewidth, zhang2015experimental, kim2008temperature}. In the first scenario, only excitons within the light cone can recombine radiatively, and the width of the light cone defines the exciton linewidth ($\Delta \mathcal{E}$). The temperature-dependent radiative decay time can then be expressed as $\tau (T) = \tau (T=0~K)/r(T)$, where $r(T)$ is the fraction of excitons occupying states within the energy window $\Delta \mathcal{E}$. With increasing temperature, the average exciton kinetic energy increases, which reduces the $r(T)$ and consequently leads to an increase in the radiative decay time. In the case of CrPS$_4$, however, $\Delta \mathcal{E}$ for the spin-allowed excitons is significantly larger than the thermal energy, implying that the majority of excitons reside within the light cone. This rules out momentum-space redistribution of the excitons as the origin of the observed increase in decay time. The second scenario requires the presence of a dark state with a longer (shorter) decay time located on the high-energy (low-energy) side of the bright exciton. However, a dark state with a longer decay time situated at higher energy would reduce the integrated PL intensity of the bright exciton~\cite{kim2008temperature}, which contradicts the observed PL enhancement and can therefore be excluded. If the dark state with a shorter decay time lies below the bright exciton, the effective decay time would be governed by the rapid recombination through the dark state. An increase in temperature enhances the thermal population of optically bright exciton states, leading to a concomitant increase in the PL intensity and a longer effective decay time. A dark state located approximately 4-10~meV below the X$_1$ exciton (\textit{SM}, Fig.~S4~\cite{SuppInfo}) could, in principle, account for the observed temperature dependence of both the integrated PL intensity and the decay time of the X$_1$ exciton. Thermal population of the higher-energy F$_0$ and X$_2$ excitons from such a dark state would require significantly larger activation energies and should therefore occur at elevated temperatures. Instead, all low-energy excitons near 1.4~eV exhibit nearly identical temperature-dependent behavior in the low-temperature regime, which requires the presence of a distinct dark state below each bright exciton. An equivalent interpretation could be that these excitons are localized by potential fluctuations in the magnetically ordered phase~\cite{dixit2014versatile, van1967optical, lao2018luminescence}. At low temperatures, excitons trapped in such potentials undergo fast non-radiative recombination, resulting in short decay times. With increasing temperature, excitons thermally escape into free-exciton states that are optically active and possess significantly longer radiative lifetimes. A common trapping potential thus naturally explains the nearly identical enhancement of PL intensity and increase in decay time observed for all low-energy excitons. A distribution of trap depths and spatially heterogeneous population exchange between trap and bright states can give rise to a stretched exponential decay ($\beta <1$), consistent with the observation.

\subsection{Excitons in the Tanabe–Sugano Framework}

\begin{figure}[htp]
\centering
\includegraphics[width=8.4cm]{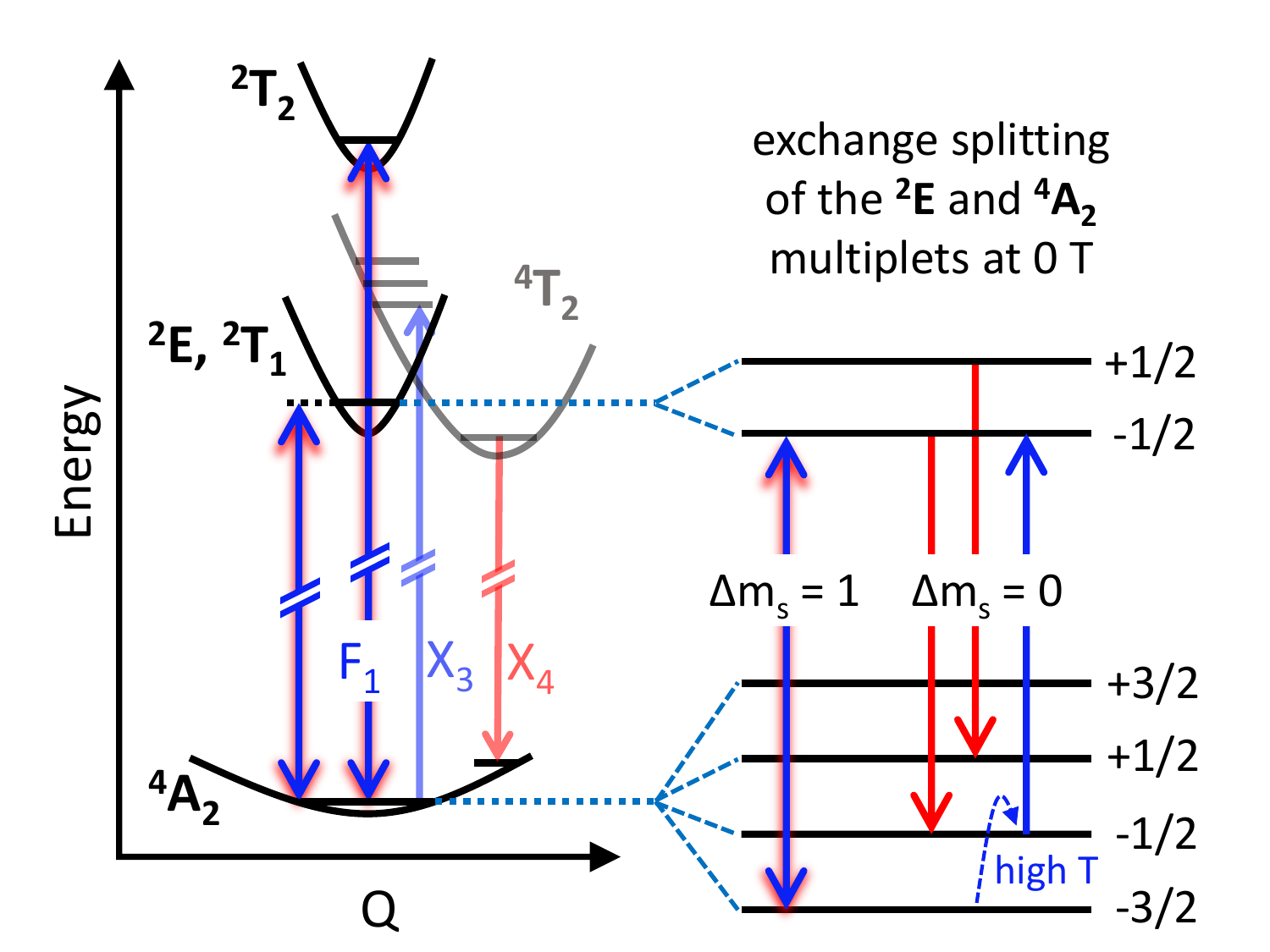}
\caption{Schematic configuration-coordinate diagram for the d$^3$ electronic configuration within the Tanabe–Sugano framework. The extended energy scheme, considering exchange splitting of the $^2E$ and $^4A_2$ states, is shown on the right. The electronic excitations (emissions) are shown by blue (red) vertical arrows.}
\label{fig:Fig5}
\end{figure}

The excitonic transitions discussed above are predominantly localized on the Cr$^{3+}$ $d$-orbitals and are therefore only weakly influenced by the extended band structure of CrPS$_4$. Owing to their localized nature, these excitations can be described within the Tanabe–Sugano framework, which applies to localized transition-metal ions embedded in a crystal-field environment. Figure~\ref{fig:Fig5} illustrates the potential-energy surfaces for a $d^3$ electronic configuration in a configuration-coordinate (CC) diagram~\cite{Tanabe1954}. The low-energy excitons near 1.4~eV are assigned to the $^2E,~^2T_1~\rightarrow~^4A_2$ transitions, while the higher-energy exciton at 1.97~eV is attributed to the $^2T_2~\rightarrow~^4A_2$ transition. The ground ($^4A_2$) and excited ($^2E,~^2T_1$, and $^2T_2$) states possess nearly identical equilibrium lattice configurations, resulting in optical absorption and emission occurring at essentially the same energy, as reflected by the coincident peak positions in the PL and PLE spectra. The $^4T_2$ excited state exhibits a significantly different equilibrium lattice configuration compared to the $^4A_2$ ground state. Consequently, optical excitation into this state occurs at higher energies, giving rise to the X$_3$ feature, whereas multi-phonon-assisted recombination occurs at substantially lower energy, producing the X$_4$ PL feature (\textit{SM}, Fig.~S4~\cite{SuppInfo}).

The presence of multiple spin-allowed and spin-flip transitions near ~1.4 eV at zero magnetic field, along with their distinct energies, however, cannot be adequately described by this simple Tanabe–Sugano framework. It requires accounting for exchange splitting in both the ground ($^4A_2$) and excited ($^2E$, $^2T_1$) states~\cite{van1967optical,wickersheim1963optically}, as schematically illustrated in Fig.~\ref{fig:Fig5}. Transitions from the $m'_s=\pm 1/2$ levels of the $^2E$ and $^2T_1$ states to the $m_s=\pm 3/2$ levels of the $^4A_2$ state give rise to spin-flip emissions ($\Delta m_s=\pm 1$), whereas transitions to the $m_s=\pm 1/2$ levels of the $^4A_2$ state correspond to spin-allowed emissions ($\Delta m_s=\pm 0$). At 5~K, the lowest-lying $m_s=\pm 3/2$ levels of the $^4A_2$ ground state (with opposite signs corresponding to the two antiferromagnetic sublattices) are predominantly populated. Consequently, the spin-allowed excitations are strongly suppressed, resulting in weak signatures of the X$_1$, F$_0$, and X$_2$ features in the 5~K PLE spectrum. As the temperature increases, the thermal population of the higher-energy sublevels of the $^4A_2$ ground state becomes significant, enhancing the intensity of these transitions in the PLE spectra, as shown in Fig.~\ref{fig:Fig2}a. The PL intensity, however, does not depend on the thermal population of the $^4A_2$ levels, and consequently the X$_1$ transition remains bright even at the lowest temperatures.

\section{Conclusions}

In summary, the bandgap and complex excitonic structure of CrPS$_4$ have been investigated using both computational and experimental approaches. CrPS$_4$ is established as a direct-gap semiconductor with a bandgap of 2.48~eV in the antiferromagnetic phase. The dominant sub-bandgap excitonic transitions, previously modeled as $^2E,~^2T_1$, and $^2T_2 \rightarrow~^4A_2$ transitions, are distinctly classified as spin-allowed and spin-flip transitions between the exchange-split levels of these states. Temperature- and magnetic field-dependent measurements reveal characteristic energy and intensity shifts of these excitons, indicating magnetic-order-induced exciton localization, redistribution among the localized $d$-states, and providing clear optical signatures of magnetic phase transitions. These results demonstrate the potential of excitons as sensitive probes of magnetic order in layered van der Waals antiferromagnets.

\section{Methods}

\textbf{Preparation of the samples.} Bulk CrPS$_4$ crystals from two sources were used: commercially available crystals from HQ Graphene and samples synthesized via the chemical vapor transport (CVT) method in a quartz ampoule. For CVT synthesis, high-purity chromium (99.99+$\%$, -60 mesh, Chemsavers, USA), phosphorus (99.9999$\%$, 1–6 mm, Wuhan Xinrong New Material Co., Ltd, China), and sulfur (99.9999$\%$, 1–6 mm, Wuhan Xinrong New Material Co., Ltd, China) were used as starting materials, with iodine (99.999$\%$, Merck, Germany) serving as the transport medium. The elements were combined in stoichiometric proportions corresponding to a total mass of 40 g and loaded into a quartz ampoule (50 × 250~mm) with iodine at a concentration of 2~mg/mL. The ampoule was evacuated and melt-sealed under high vacuum. The sealed ampoule was initially heated in a muffle furnace following a stepwise temperature profile: 450~$^{\circ}$C for 25~h, 500 ~$^{\circ}$C for 50~h, 600 ~$^{\circ}$C for 50~h, and 700 ~$^{\circ}$C for 50~h, with heating and cooling rates of 1~$^{\circ}$C/min. The resulting polycrystalline CrPS$_4$ and residual elements were then placed in a horizontal two-zone furnace. A reverse temperature gradient was applied initially (source zone 600~$^{\circ}$C, growth zone 750~$^{\circ}$C) for 50 h, followed by a standard gradient with the source zone at 750~$^{\circ}$C and the growth zone decreasing from 700~$^{\circ}$C to 650~$^{\circ}$C over 7 days. For the subsequent 7 days, a thermal gradient of 100~$^{\circ}$C was maintained. Finally, the growth zone was held at 400~$^{\circ}$C and the source zone at 100~$^{\circ}$C to remove residual transport medium. Crystals with typical sizes of 5–15~mm were recovered from the ampoule inside an argon-filled glovebox, yielding approximately 40 g of CrPS$_4$. The crystals were mounted on SiO$_2$/Si substrates either using adhesive for thicker samples or via mechanical exfoliation for thinner flakes. Sample thicknesses were chosen to minimize interference effects in PL, PLE, and absorption measurements.  

\textbf{Computational methodology:} The $QSG\hat{W}$~\cite{Cunningham2023} calculations for the bulk antiferromagnetic (AFM) and ferromagnetic (FM) phases were performed with magnetic two-layer unit cell (containing 24 atoms) structure by applying periodic boundary conditions along all three directions. Every layer contains 12 atoms, where Cr atoms are ferromganetically aligned, while the inter-layer coupling is antiferromagnetic. The static quasi-particle self-energy $\Sigma^0(\mathbf{k})$ is generated on a 10$\times$10$\times$4 k-mesh, and the dynamical self-energy ($\Sigma(\mathbf{k})$) is generated on a 5$\times$5$\times$2 k-mesh for the 24 atom cell. The convergence threshold for the self-energy is set at $1{\cdot}10^{-5}$ Ry while for charge it is set at $1{\cdot}10^{-6}$ Ry. The paramagnetic (PM) calculations are performed by extending super-cells to 48 atoms, where the cells are primarily enhanced in the $ab$ plane, while keeping the cell length along the $c$ axis fixed. QS$GW$ calculations are performed involving all occupied and unoccupied states for all AFM, FM and PM cells. Bethe-Salpeter equation (BSE) is solved for the $QSG\hat{W}$ Hamiltonian for the 24-atom AFM and FM cells, including 36 valence bands and 16 conduction bands with spin-orbit coupling. As cells are doubled for the PM simulations, the number of BSE active states is proportionally increased. To simulate the non-local spin disorder for the bulk PM phase, 48 atoms were included in the supercell and 48 interstitial sites were also added to augment the basis with floating orbitals. For the PM phase, local spin orientations were arranged in a quasi-random configuration to minimize the difference between the quasirandom and true random site correlation functions. An objective function composed from 420 pair and 324 triplet functions was minimized, following the approach by Zunger et al.~\cite{zunger}.

For simulating site-local spin disorder and all triplet and singlet atomic multiplets, dynamical mean-field theory (DMFT) calculations were performed. Within our QSGW+DMFT calculations for CrPS$_{4}$, we projected the lattice problem on the Cr d orbitals following the prescription of Haule. The five 3$d$ orbitals of the transition metal atom are included in the correlated Anderson impurity, while states within +10 and -10 eV around the Fermi energy are included in the bath. For Cr, Hubbard parameters are $U=3.2$~eV, $J=0.7$~eV, and a fully localized limit double counting correction is used to get the band gaps consistent with the parameter-free estimations from of $\mathrm{QSG\hat{W}}$. Once the band gaps are synchronized between the two methods, the higher-order charge-charge correlators are computed ~\cite{acharyaTheoryColorsStrongly2023} from the exact diagonalization solver. To single out the correlated subspace, a procedure of embedding, originally introduced by Haule in the LAPW basis of the Wien2k package, is developed in the Full-Potential Linear Muffin-Tin Orbitals~\cite{methfessel2000full}. The technical information on the embedding process, choices of hybridization window, and the constrained RPA calculations performed to choose the Hubbard parameters are discussed in our previous work~\cite{PhysRevB.95.041112}.

 \textbf{Optical measurements} Micro-optical setups were employed for PL and PLE measurements. For PL, the samples were excited using a continuous-wave 515.5~nm laser focused through a 50$\times$ microscope objective, producing a spot diameter of approximately 1~$\mu$m. The emitted luminescence was collected by the same objective, dispersed by a 0.7~m focal length spectrometer, and detected using a silicon-based charge-coupled device (CCD) camera cooled to 120~K. The same setup was used for PLE measurements, with the 515.5~nm laser replaced by a supercontinuum tunable laser. Appropriate short- and long-pass filters were employed to clean the excitation wavelength and the collected luminescence. The PLE excitation power was maintained in the tens of $\mu$W range. Time-resolved PL measurements were performed by using a single-photon counter (Excelitas SPCM-AQRH-16-BR1) triggered by laser pulses. Excitation was provided by a 532~nm picosecond laser (VisUV, PicoQuant). Spectral filtering was performed using a linear polarizer in combination with bandpass filters: a 980~nm bandpass filter, angle-tuned to access low-energy excitons, and a 633~nm bandpass filter for high-energy excitons. Temperature- and magnetic-field-dependent experiments were performed by mounting the sample on an x–y–z piezo stage within a free-beam insert positioned inside a superconducting magnet, allowing magnetic fields up to 14~T and tunable sample temperatures.

\section{Acknowledgments}
 This project was supported by the Ministry of Education (Singapore) through the Research Centre of Excellence program (grant EDUN C-33-18-279-V12, I-FIM), and Academic Research Fund Tier 2 (MOE-T2EP50122-0012). This material is based upon work supported by the Air Force Office of Scientific Research and the Office of Naval Research Global under award number FA8655-21–1-7026. This work was authored in part by the National Laboratory of the Rockies for the U.S. Department of Energy (DOE) under Contract No. DE-AC36-08GO28308. Funding was provided by the Computational Chemical Sciences program within the Office of Basic Energy Sciences, U.S. Department of Energy.  SA, DP and MvS acknowledge the use of the National Energy Research Scientific Computing Center, under Contract No. DE-AC02-05CH11231 using NERSC award BES-ERCAP0021783, and also acknowledge that a portion of the research was performed using computational resources sponsored by the Department of Energy's Office of Energy Efficiency and Renewable Energy and located at the National Laboratory of the Rockies, and computational resources provided by the Oakridge leadership Computing Facility. The views expressed in the article do not necessarily represent the views of the DOE or the U.S. Government. The U.S. Government retains and the publisher, by accepting the article for publication, acknowledges that the U.S. Government retains a nonexclusive, paid-up, irrevocable, worldwide license to publish or reproduce the published form of this work, or allow others to do so, for U.S. Government purposes. M.P. acknowledges support from the CENTERA2, FENG.02.01-IP.05- T004/23 project funded within the IRA program of the FNP Poland, cofinanced by the EU FENG Programme and from the ERC-AG TERAPLASM (No. 101053716) project. C.F. acknowledges support from ANR-23-QUAC-0004 and from the CEFIPRA project (No.~7104-2). Z.S. acknowledges support from the Ministry of Education, Youth and Sports (MEYS) under project LUAUS25268, the project Advanced Functional Nanorobots (reg.~No.~CZ.02.1.01/0.0/0.0/15\_003/0000444 financed by the EFRR), and the ERC-CZ programme (project LL2101) from MEYS.


\providecommand{\noopsort}[1]{}\providecommand{\singleletter}[1]{#1}%

\newpage
\pagenumbering{gobble}

\begin{figure}[htp]
\includegraphics[page=1,trim = 18mm 18mm 18mm 18mm,
width=1.0\textwidth,height=1.0\textheight]{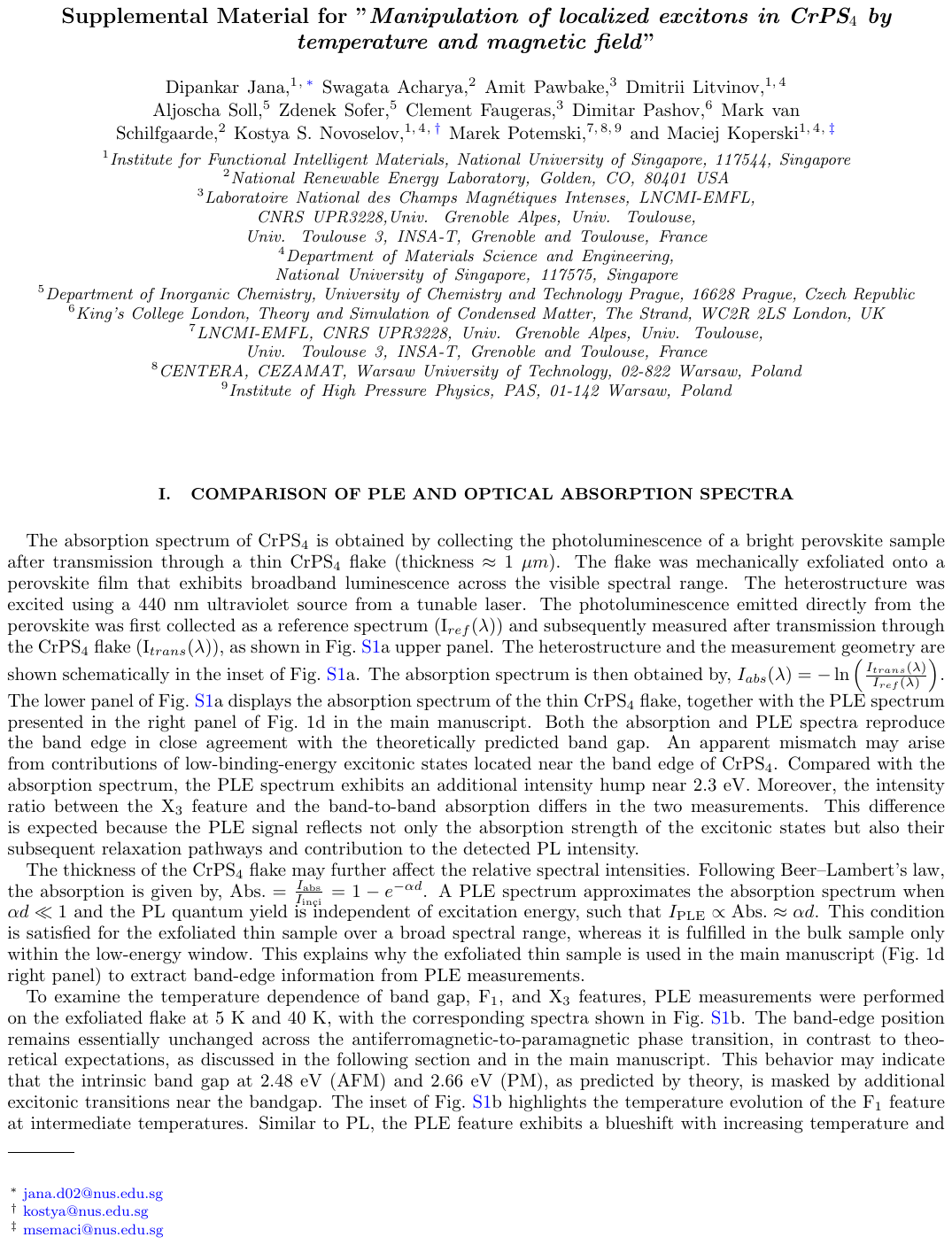}
\end{figure}

\newpage

\begin{figure}[htp]
   \includegraphics[page=2,trim = 18mm 18mm 18mm 18mm,
width=1.0\textwidth,height=1.0\textheight]{Spin_allowed_excitons_in_CrPS4.pdf}
\end{figure}
\newpage

\begin{figure}[htp]
   \includegraphics[page=3,trim = 18mm 18mm 18mm 18mm,
width=1.0\textwidth,height=1.0\textheight]{Spin_allowed_excitons_in_CrPS4.pdf}
\end{figure}

\begin{figure}[htp]
   \includegraphics[page=4,trim = 18mm 18mm 18mm 18mm,
width=1.0\textwidth,height=1.0\textheight]{Spin_allowed_excitons_in_CrPS4.pdf}
\end{figure}

\newpage

\begin{figure}[htp]
   \includegraphics[page=5,trim = 18mm 18mm 18mm 18mm,
width=1.0\textwidth,height=1.0\textheight]{Spin_allowed_excitons_in_CrPS4.pdf}
\end{figure}
\newpage

\begin{figure}[htp]
   \includegraphics[page=6,trim = 18mm 18mm 18mm 18mm,
width=1.0\textwidth,height=1.0\textheight]{Spin_allowed_excitons_in_CrPS4.pdf}
\end{figure}

\begin{figure}[htp]
   \includegraphics[page=7,trim = 18mm 18mm 18mm 18mm,
width=1.0\textwidth,height=1.0\textheight]{Spin_allowed_excitons_in_CrPS4.pdf}
\end{figure}

\end{document}